\documentclass[twocolumn,showpacs,preprintnumbers,amsmath,amssymb,prl]{revtex4}

\usepackage{graphicx}
\usepackage{dcolumn}
\usepackage{bm}

\usepackage{mathptmx, courier, pifont}
\usepackage[scaled=0.92]{helvet}
\usepackage[T1]{fontenc}
\usepackage{textcomp}

\newcommand{\quarterthin}{\kern 0.0417em}

\begin{document}


\title{Coulomb energy difference as a probe of isospin-symmetry breaking
in the upper $fp$-shell nuclei}

\author{K.~Kaneko$^{1}$, T.~Mizusaki$^{2}$,
 Y.~Sun$^{3,4}$, S.~Tazaki$^{5}$, G.~de~Angelis$^{6}$ }

\affiliation{
$^{1}$Department of Physics, Kyushu Sangyo University, Fukuoka
813-8503, Japan \\
$^{2}$Institute of Natural Sciences, Senshu University, Tokyo
101-8425, Japan \\
$^{3}$Department of Physics, Shanghai Jiao Tong
University, Shanghai 200240, People's Republic of China \\
$^{4}$Institute of Modern Physics, Chinese Academy of Sciences,
Lanzhou 730000, People's Republic of China \\
$^{5}$Department of Applied Physics, Fukuoka University, Fukuoka
814-0180, Japan \\
$^{6}$Laboratori Nazionali di Legnaro dell'INFN, Legnaro (Padova),
I-35020, Italy
}

\date{\today}

\begin{abstract}

The anomaly in Coulomb energy differences (CED) between the isospin
$T=1$ states in the odd-odd $N=Z$ nucleus $^{70}$Br and the analogue
states in its even-even partner $^{70}$Se has remained a puzzle.
This is a direct manifestation of isospin-symmetry breaking in
effective nuclear interactions. Here, we perform large-scale
shell-model calculations for nuclei with $A=66-78$ using the new
filter diagonalization method based on the Sakurai-Sugiura
algorithm. The calculations reproduce well the experimental CED. The
observed negative CED for $A=70$ are accounted for by the
cross-shell neutron excitations from the $fp$-shell to the $g_{9/2}$
intruder orbit with the enhanced electromagnetic spin-orbit
contribution at this special nucleon number.

\end{abstract}


\maketitle


Isospin is a fundamental concept in nuclear and particle physics.
The isospin symmetry was introduced under the assumption of charge
independence of the nuclear force \cite{Wigner37}. Historically, the
study of this symmetry led directly to the discovery and
understanding of quarks. However, it is well known that this
symmetry is only approximate because of the existence of the Coulomb
interaction and isospin-breaking interactions among nucleons,
leading to small differences, for example, in the binding energy of
mirror-pair nuclei and in the excitation energy of the same spin,
$J$, between isobaric analogue states (IAS) of the same isospin,
$T$.

A nucleus is a quantum many-body system with finite size, which
generally shows two unique features in structure: the shell effect
with the presence of strong spin-orbit interaction
\cite{ShellStructure} and the nuclear deformation associated with
collective motion \cite{BohrMottelson}. To properly describe these
aspects in the framework of nuclear shell models, effective
interactions must be involved. Thus the effects of isospin-symmetry
breaking can manifest themselves through structure changes in the
vicinity of the $N=Z$ line, providing information on the
$T_z$-dependence of the effective interactions. The effects have
been extensively studied for nuclei in the upper $sd$- and the lower
$fp$-shell regions (see Ref. \cite{Bentley07} for review), where a
remarkable agreement between experimental mirror energy differences
(MED) and shell-model calculations has been found, allowing a clear
identification of the origin of isospin-symmetry breaking in
effective nuclear interactions.

The Coulomb energy difference (CED), defined by
\begin{equation}
 {\rm CED}({J}) = E_{x}(J,T=1,T_{z}=0) - E_{x}(J,T=1,T_{z}=1),
  \label{eq:0}
\end{equation}
is often regarded as a measure of isospin-symmetry breaking in
effective nuclear interactions which include the Coulomb force
\cite{Nara07,Angelis01}. In Eq. (\ref{eq:0}), $E_{x}(J,T,T_{z})$
are the excitation energies of IAS with spin $J$ and isospin $T$,
distinguished by different $T_{z}$ (the projection of $T$).

\begin{figure}[t]
\includegraphics[totalheight=6.0cm]{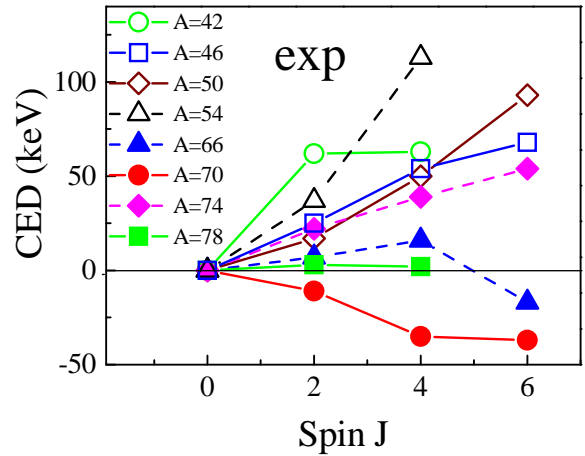}
  \caption{(Color online) Experimental CED between the isospin $T=1$ states
in odd-odd $N=Z$ nuclei and the IAS in even-even nuclei for mass
numbers $A=42-78$. Data were taken from Refs.
\cite{Bentley07,Angelis01,Nara07,Angelis12}.}
  \label{fig1}
\end{figure}

More complex shell structures are expected for heavier mass regions.
In the upper $fp$-shell, for example, abrupt structure changes along
the $N=Z$ line \cite{Hasegawa07} and the phenomenon of shape
coexistence \cite{France03,Fischer00,Gade05} are known. Being pushed
down to the lower shell by the spin-orbit interaction
\cite{ShellStructure}, the $g_{9/2}$ intruder orbit and its
interplay with the $fp$-shell orbits play a key role in the overall
structure. However, the influence of these structure changes on CED
has not been explored. In Fig. \ref{fig1}, we show the experimental
CED for nuclei with mass numbers $A=42$ to 78, together with the
very recent data of the $4_{1}^{+}$ and $6_{1}^{+}$ states in $A=$
66 \cite{Angelis12}. One observes that, while most of the CED values
are positive, only the mass number $A=$ 70 has negative CED
\cite{Angelis01,Jenkins02}. There have been suggestions to explain
such an anomalous behavior but without much success. In Ref.
\cite{Angelis01}, it was suggested that this behavior could be
attributed to the Thomas-Ehrman effect \cite{Thomas52,Ehrman51} of
the loosely bound proton in $^{70}$Br. However, as pointed out in
Ref. \cite{Nara07}, this suggestion cannot explain the observed
positive CED in other systems such as $A=$ 66, 74, 78 because these
systems have similar binding energy differences as in $A=$ 70. In
Ref. \cite{Nara07}, it was suggested that the anomaly may be
qualitatively accounted for by small variations in the Coulomb
energy due to shape differences in the pair of nuclei. The authors
in Ref. \cite{Nara07} tried to argue that such deformation changes
occur only for a prolate shape, and not for transition from an
oblate ground state to excited configurations \cite{Nara07}.
However, recent measurements and theoretical calculations indicate
that both $^{70}$Br and $^{70}$Se are associated with an oblate
shape \cite{Jenkins02,Ljungvall08}. Most recently, the anomalous
behavior has been suggested as due to different mixing of competing
shapes within the members of the isobaric multiplet
\cite{Angelis12}.

In the past few years, experimental data for mirror nuclei above the
doubly magic $^{56}$Ni have become available. The MED in the $A\sim 60$
mass region were discussed \cite{Ekman05,Andersson05}. It was reported
that the contribution to MED from the electromagnetic spin-orbit
term $\varepsilon_{ls}$ is significant, while the monopole Coulomb
term $\varepsilon_{ll}$ is not. Here, detailed variations in nuclear
shell structure enter into the discussion. Energy shifts due to the
$\varepsilon_{ls}$ term increase the gap between the $p_{3/2}$,
$f_{5/2}$ and the $g_{9/2}$ orbit for neutrons but reduce that for
protons. As a consequence, excitations involving these orbits
contribute differently to MED. We have recently investigated MED
between the isobaric analogue states in the mirror-pair nuclei
$^{67}$Se and $^{67}$As using large-scale shell-model calculations
\cite{Kaneko10}. The calculations reproduced well the experimental
MED \cite{Orlandi09}, and suggested that the $\varepsilon_{ls}$ term
provides the main contribution to the observed MED values. The
negative MED values observed in the $A$ = 67 mirror nuclei were
explained by the single-particle-energy shifts due to the
$\varepsilon_{ls}$ contribution. We may thus expect that the
$\varepsilon_{ls}$ term plays a similarly important role for the
anomaly in CED in the $A=$ 70 IAS.

To study CED and probe isospin-symmetry breaking in finite nuclei,
calculations for energy levels with an accuracy of a few keV are
required. The current state-of-the-art shell-model calculations with
the m-scheme using the Lanczos diagonalization method \cite{Lanczos}
can handle a matrix dimension of the order of 10$^{10}$. However, it
is difficult for the Lanczos method to obtain solutions for $T\sim
1$ states in odd-odd $N=Z$ nuclei under the isospin-symmetry
breaking with the Coulomb interaction. As for instance, it is
impossible to obtain the $T\sim 1$, $J=4$ state because of the
presence of numerous $T\sim 0$ states below it. To solve this
problem, a breakthrough in shell-model calculations is needed. Quite
recently, the filter diagonalization method for large-scale
shell-model calculations has been proposed by some of us
\cite{Mizusaki10}, which is based on a new algorithm for
diagonalization suggested by Sakurai and Sugiura \cite{SS03} (called
the SS method). The filter diagonalization procedure can overcome
the difficulty mentioned above \cite{Mizusaki11}.

In this Letter, shell-model calculations are performed by using the
filter diagonalization method in the $pf_{5/2}g_{9/2}$ model
space. We employ the recently-proposed JUN45 interaction
\cite{Honma09}, a realistic effective interaction based on the
Bonn-C potential and adjusted to the experimental data of the $A=63
\sim 96$ mass region. We try to solve a general shell-model
eigenvalue equation
\begin{equation}
\hat{H}|\Phi_{k}\rangle =e_{k}|\Phi_{k}\rangle,
\label{eq.2}
\end{equation}
where $\hat{H}$ is the shell-model Hamiltonian, and $e_{k}$ and
$|\Phi_{k}\rangle$ are eigenvalues and eigenfunctions, respectively.
The spirit of the filter diagonalization method \cite{Mizusaki10} is
that in order to reduce the large-scale eigenvalue problem to a
smaller one, one introduces moments $\mu_{p}(p=0,1,2,\cdots)$
defined by the Cauchy's integral
\begin{equation}
\displaystyle \mu_{p}=\frac{1}{2\pi i}\int_{\Gamma}\langle\psi|
\frac{(z-\varepsilon)^{p}}{z-\hat{H}}|\phi\rangle dz,
\label{moment}
\end{equation}
where $|\psi\rangle$ and $|\phi\rangle$ are arbitrary wave functions,
$\varepsilon$ is a given energy for a target region, and $\Gamma$
is an integration contour. Following the SS method \cite{SS03}, one can
extract the eigenvalues $e_{k}(k\in\Gamma)$ inside the closed curve
$\Gamma$ from these moments. In this way, the large-scale eigenvalue
problem inside $\Gamma$ is reduced to
\begin{equation}
Mx=\lambda Nx,
\label{ediag}
\end{equation}
where $M$ and $N$ are $n\times n$ Hankel matrices expressed by the
moments $\mu_{p}$ and the eigenvalues $\lambda_{0},\cdots ,
\lambda_{n-1}$ in the interior of $\Gamma$. After diagonalizing
(\ref{ediag}), one obtains the eigenvalues
$e_{k}=\varepsilon+\lambda_{k}$ inside the integration contour
$\Gamma$ (Details were discussed in Ref. \cite{Mizusaki10}). For the
present case, we apply filter diagonalizations by taking the
$T_{-}|\psi_{J}^{T=1}\rangle$ state as $|\psi\rangle$ and
$|\phi\rangle$ in Eq. (\ref{moment}) for the odd-odd $N=Z$ nucleus,
where $T_{-}$ is the lowering operator of isospin and
$|\psi_{J}^{T=1}\rangle$ is an isovector $T=1$ state with spin $J$
in the even-even $N=Z+2$ nucleus. Therefore
$T_{-}|\psi_{J}^{T=1}\rangle$ is a good approximation to the wave
function of the CED state. In sharp contrast to the usual Lanczos
method, by taking $\Gamma$ around the energy of this state, the
filter diagonalization handles only the target CED state. This is
the reason why the filter diagonalization can succeed in the
calculation.

CED have been studied in the shell-model framework by using the
formalism introduced by Zuker {\it et al.} \cite{Zuker02}. In this
description, the Coulomb interaction is separated into a monopole
term $V_{Cm}$ and a multipole term $V_{CM}$. While $V_{Cm}$ accounts
for single-particle and bulk effects, $V_{CM}$ contains all the
rest. The monopole term $V_{Cm}$ is further divided into the
single-particle correction $\varepsilon_{ll}$, the radial term
$V_{Cr}$, and the electromagnetic spin-orbit term
$\varepsilon_{ls}$. It has been reported that the radial term
$V_{Cr}$ does not give a correct description in the lower part of
the upper $fp$-shell \cite{Ekman05}, and furthermore, this term may
be important only for high-spin states \cite{Kaneko10}. Therefore,
$V_{Cr}$ is not considered in the present calculation. The isospin
nonconserving interaction is also neglected for the upper half of
the $fp$-shell region because the $f_{7/2}$ orbit is almost not
active.

The single-particle energy shift $\varepsilon_{ls}$ takes into
account the relativistic spin-orbit interaction \cite{Nolen69}. This
interaction originates from the Larmor precession of nucleons in
electric fields due to their magnetic moments, which, as well known,
affects the single-particle energy spectrum. $\varepsilon_{ls}$ can
be written as \cite{Nolen69}
\begin{eqnarray}
 \varepsilon_{ls} & = & (g_{s}-g_{l})\frac{1}{2m_{N}^{2}c^{2}}\left( \frac{1}{r}\frac{dV_{c}}{dr}
  \right)\langle \hat{l}\cdot\hat{s}\rangle,
          \label{eq:2}
\end{eqnarray}
where $m_{N}$ is the nucleon mass and $V_{c}$ is the Coulomb potential
due to the $^{56}$Ni core. The free values of the
gyromagnetic factors, $g_{s}^{\pi}=5.586$, $g_{l}^{\pi}=1$ for
protons and $g_{s}^{\nu}=-3.828$, $g_{l}^{\nu}=0$ for neutrons, are
used. By assuming a uniformly charged sphere, $\varepsilon_{ls}$ is
calculated in the present work by using the harmonic-oscillator
single-particle wave functions. Depending on proton or neutron
orbits, the shift can have opposite signs. It depends also on the
spin-orbit coupling, as for instance $\langle
\hat{l}\cdot\hat{s}\rangle =l/2$ when $j=l+s$ and $\langle
\hat{l}\cdot\hat{s}\rangle =-(l+1)/2$ when $j=l-s$.  As one will
see, these strongly affect the results in the following discussion.

\begin{figure}[t]
\includegraphics[totalheight=5.5cm]{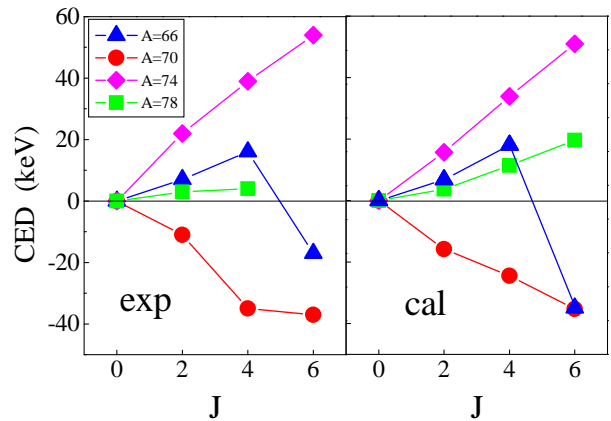}
  \caption{(Color online) Comparison of calculated CED with experimental data
for mass number $A=$66, 70, 74, and 78.  Note that for $A=$66, the
calculated CED correspond to the second 6$^{+}$ states.}
  \label{fig2}
\end{figure}

With inclusion of the $V_{CM}$, $\varepsilon_{ll}$, and
$\varepsilon_{ls}$ terms, shell-model calculations using the filter
diagonalization method are carried out in the $pf_{5/2}g_{9/2}$
model space for odd-odd $N=Z$ nuclei and their even-even IAS
partners for $A=$66, 70, 74, and 78. In Fig. \ref{fig2}, the
calculated CED are compared with available experimental data. As can
be seen, the calculation reproduces the experimental CED remarkably
well. In particular, the negative CED for $A=70$ and the large,
positive CED for $A=74$ are correctly obtained.

\begin{figure}[b]
\includegraphics[totalheight=7.0cm]{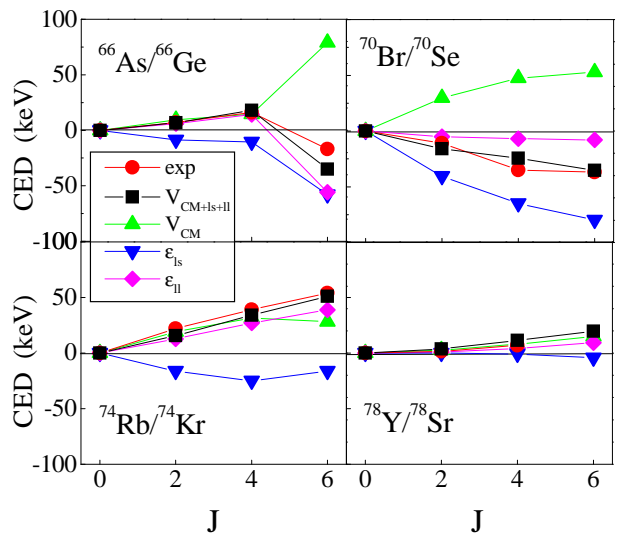}
  \caption{(Color online) Decomposition of theoretical CED for states
shown in Fig. \ref{fig2} into three terms (see text for
explanation). The total CED are denoted by black square symbols.}
  \label{fig3}
\end{figure}

We now analyze the results by looking at the components. Figure
\ref{fig3} shows the total CED denoted by $V_{CM+ls+ll}$, and the
separated multipole, spin-orbit, and orbital parts by $V_{CM}$,
$\varepsilon_{ls}$, and $\varepsilon_{ll}$, respectively. For the
$A=66$ pair $^{66}$As/$^{66}$Ge, the spin-orbit component
$\varepsilon_{ls}$ has a negative value for the $J=0$, 2 and 4
states, while the other two, $V_{CM}$ and $\varepsilon_{ll}$, are
positive. Since the positive values cancel out the negative ones,
the net CED are small and positive. At $J=6$, the sudden change from
this behavior can be understood as the fact that the involving
states are not the first, but the second $6^+$ states in the $A=66$
pair. The values for the $A=70$ pair $^{70}$Br/$^{70}$Se are large
and negative for $\varepsilon_{ls}$, but positive for $V_{CM}$ and
nearly zero for $\varepsilon_{ll}$. Since the absolute values of
$\varepsilon_{ls}$ are larger, the total CED are therefore negative.
This suggests that the spin-orbit contribution is responsible for
the observed negative CED in $^{70}$Br/$^{70}$Se. For the $A=74$
pair $^{74}$Rb/$^{74}$Kr, the components indicate a similar overall
behavior as those of the lower spin states in the $A=66$ pair.
However, both $V_{CM}$ and $\varepsilon_{ll}$ are found larger in
the $A=74$ pair, and thus are dominant in the summation. Therefore,
the $A=74$ CED are large and positive. For $^{78}$Y/$^{78}$Sr, all
components are small, and the total CED therefore indicate small and
positive values.


Our calculation suggests that for all the pairs studied, the
spin-orbit component, $\varepsilon_{ls}$, is negative. The effect is
particularly enhanced for $A=70$. Thus the spin-orbit term appears
to be the origin of the observed CED anomaly in $^{70}$Br/$^{70}$Se.
But why the absolute values of the spin-orbit component are
anomalously large only for $^{70}$Br/$^{70}$Se? We previously
mentioned that the spin-orbit term affects single-particle
corrections differently for neutrons and protons. We may discuss
these with the following estimate. Due to this interaction, the
proton $g_{9/2}$ orbit is lowered by about 66 keV, while the
$f_{5/2}$ orbit is raised by about 66 keV. Then the energy gap
between the proton $g_{9/2}$ and $f_{5/2}$ orbit decreases roughly
by 132 keV. On the contrary, the neutron $g_{9/2}$ orbit is raised
by about 55 keV and the $f_{5/2}$ orbit is lowered by about 55 keV,
and therefore, the spin-orbit contribution enhances the neutron
shell gap between the $g_{9/2}$ and $f_{5/2}$ orbit roughly by 110
keV. The total effect thus amplifies the difference between the
neutron and proton orbits by about 242 keV.

By introducing the difference in neutron occupation number between
the ground state of $0_{1}^{+}$ and the excited state with spin $J$
\begin{eqnarray}
 \Delta n_{g_{9/2}}^{J} & = & n_{g_{9/2}}^{J} - n_{g_{9/2}}^{J=0},
 \label{eq:3}
\end{eqnarray}
it is straightforward to see that the excited states in $^{70}$Se
lie higher than those in $^{70}$Br with the same spin, and
therefore, CED in Eq. (\ref{eq:0}) become negative. Since the energy
gap between the $g_{9/2}$ and $fp$ shell for neutrons is by 242 keV
larger than that for protons, the spin-orbit components can be
estimated to be -33.4, -58.8, and -77.2 keV for spin $J=$ 2, 4, and 6
from $(\Delta n_{g_{9/2}}^{J}(^{70}{\rm Br}) - \Delta
n_{g_{9/2}}^{J}(^{70}{\rm Se}))\times$ 242 keV, respectively. These
values agree well with the calculated spin-orbit components for
$^{70}$Br/$^{70}$Se in Fig. \ref{fig3}.

\begin{figure}[t]
\includegraphics[totalheight=5.2cm]{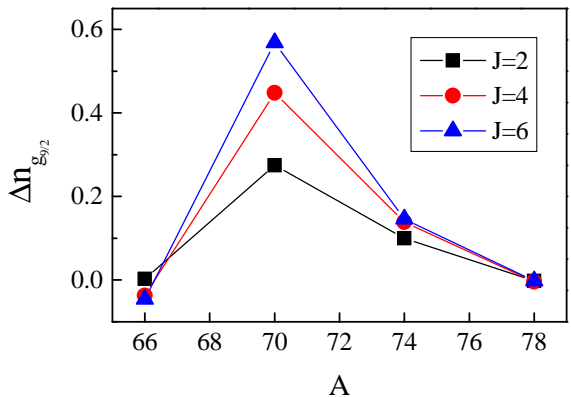}
  \caption{(Color online) Difference in
the neutron occupation number $n_{g_{9/2}}^{J}$ between the ground
state and the excited states (defined in Eq. (\ref{eq:3})) with spin
$J=$2, 4, and 6 for $^{66}$Ge, $^{70}$Se, $^{74}$Rb, and $^{78}$Y.}
  \label{fig4}
\end{figure}

Figure \ref{fig4} shows the differences in neutron occupation
number, $\Delta n_{g_{9/2}}^{J}$, between the ground $0_{1}^{+}$
state and the excited $J$-states for $A=$ 62, 66, 70, and 74. One
can see that the numbers are larger for $A=70$. The enhanced
neutron excitations from the $fp$ orbits into the $g_{9/2}$ orbit
combined with the single-particle energy shifts $\varepsilon_{ls}$
due to the spin-orbit interaction are the major factors to explain
the large negative CED in $^{70}$Br/$^{70}$Se. At the mass number
$A=70$, the increase of the $g_{9/2}$ occupation reflects the
enhanced excitations from the $fp$ shell into the $g_{9/2}$ orbit,
which means that $^{70}$Br and $^{70}$Se are transitional nuclei. In
fact, it has been known that the $N=Z$ nuclei around $A=70$ such as
$^{68}$Se and $^{72}$Kr do not correspond to a single well-developed
shape, and may exhibit the phenomenon of oblate-prolate shape
coexistence \cite{France03,Angelis12,Fischer00,Gade05}. In addition,
a shape phase transition with an abrupt change in structure when the
proton and neutron numbers cross $N=Z=35$ has been suggested
\cite{Hasegawa07}.

For $^{70}$Se, our calculation with the standard effective charges
1.5$e$ for protons and 0.5$e$ for neutrons obtains
$B(E2,I\rightarrow I-2)$ values of 345, 539, and 594 $e^{2}fm^{4}$
for $E2$ transitions de-exciting the $J^{\pi}=2_{1}^{+}, 4_{1}^{+},$
and $6_{1}^{+}$ states, respectively. These are in a good agreement
with the experimental values 342 (19), 370 (24), and 530 (96)
$e^{2}fm^{4}$ found in Ref. \cite{Ljungvall08}. Our calculated
spectroscopic quadrupole moments are 37.5, 50.2, and 55.5 $efm^{2}$,
respectively for the $J^{\pi}=2_{1}^{+}, 4_{1}^{+},$ and $6_{1}^{+}$
states. Using these values and assuming an axial deformation, we may
estimate the quadrupole deformation as $-0.21$, $-0.22$, and $-0.22$
for the $J^{\pi}=2_{1}^{+}, 4_{1}^{+},$ and $6_{1}^{+}$ states,
respectively, suggesting that the $^{70}$Se yrast states are
oblately deformed. This is in consistence with the conclusions from
the recent measurement \cite{Ljungvall08}. The deformation is
roughly constant, and does not show changes with increasing spin.
Thus from our shell-model calculation, the negative CED for $A=70$
are not attributed to subtle differences in the Coulomb energy as
shapes evolve with spin \cite{Nara07}.

In conclusion, by performing modern shell-model calculations, we
have investigated the CED effects between the isospin $T=1$ states
in odd-odd $N=Z$ nuclei and the isobaric analogue states in their
even-even neighbors for the upper $fp$-shell nuclei. In order to
obtain the $T=1$ states for odd-odd $N=Z$ nuclei, we have gone
beyond the usual Lanczos method by employing the filter
diagonalization method for the first time in application. It has
been shown that the anomalous CED found for the pair $^{70}$Br and
$^{70}$Se originates from neutron excitations from the $fp$-shell to
the $g_{9/2}$ intruder orbit reflected in a sudden enhancement in
the electromagnetic spin-orbit term for $A=70$. The study has shown
how the structure changes manifest themselves in a measure of the
isospin-symmetry breaking in effective nuclear interactions.

This work was partially supported by the National Natural Science
Foundation of China under contract No. 11075103 and 11135005.



\end{document}